\documentclass[preprint,showpacs,showkeys,preprintnumbers,amsmath,amssymb,superscriptaddress]{revtex4}
\usepackage{graphicx}
\usepackage{dcolumn}
\usepackage{bm}

\begin{document}
\title{Charged particles' tunneling from a noncommutative charged
black hole}

\author{S. Hamid Mehdipour}

\email{mehdipour@iau-lahijan.ac.ir}

\affiliation{Islamic Azad University, Lahijan Branch, P. O. Box
1616, Lahijan, Iran}

\date{\today}
\begin{abstract}
We apply the tunneling process of charged massive particles through
the quantum horizon of a Reissner-Nordstr\"{o}m black hole in a new
noncommutative gravity scenario. In this model, the tunneling
amplitude on account of noncommutativity influences in the context
of coordinate coherent states is modified. Our calculation points
out that the emission rate satisfies the first law of black hole
thermodynamics and is consistent with an underlying unitary theory.
\end{abstract}


\maketitle
\section{\label{sec:1}Introduction}
Over three decades ago, Stephan Hawking \cite{haw} found that,
utilizing the procedure of quantum field theory in curved spacetime,
the radiation spectrum is almost like that of a black body, and can
be described by a characteristic Hawking temperature with a purely
thermal spectrum which yields to non-unitarity of quantum theory
where maps a pure state to a mixed state. It has been proposed that
Hawking radiation can be extracted from the null geodesic method
suggested by Parikh and Wilczek \cite{par1}. In their method, they
take the back-reaction effects into consideration and present a
leading correction to the probability of massless particles
tunneling across the horizon. The tunneling process clarifies that
the extended radiation spectrum is not precisely thermal which leads
to unitarity. Recently, Nicolini, Smailagic and Spallucci (NSS)
\cite{nic1} derived that black hole in a new model of
noncommutativity does not allow to decay lower than a {\it minimal
mass} $M_0$, i.e. black hole remnant (see also \cite{nic2,nic2.1}) .
If we really believe the idea of stable black hole remnants due to
the fact that there are some exact continuous global symmetries in
nature \cite{bek}, and also do not find any correlations between the
tunneling rates of different modes in the black hole radiation
spectrum, then these leave only one possibility: the information
stays inside the black hole and can be retained by a stable
Planck-sized remnant \cite{ham1,ham2} {\footnote {However, in
Ref.~\cite{ham5}, we have analyzed a specific form of the
noncommutative-inspired Vaidya solution which leads to a zero
remnant mass in the long-time limit, i.e. an instable black hole
remnant.}}. Although, this issue is then accepted if information
conservation is really conserved in our universe. Thus we proceed
our work with hope that this model of noncommutativity can provide a
way to explain how the charged black hole decays, particularly in
its final stages.

The NSS model of noncommutativity of coordinates that is carried on
by the Gaussian distribution of coherent states, is consistent with
Lorentz invariance, Unitarity and UV-finiteness of quantum field
theory \cite{sma,sma.1,sma.2,sma.3,sma.4}. Since the
noncommutativity of spacetime is an innate property of the manifold
by itself even in absence of gravity, then some kind of divergences
which emerge in general relativity and black hole physics, can be
removed by it. Then with hope to cure the divergences of evaporation
process of black hole physics we apply both back-reaction and
noncommutativity effects to proceed the radiative process. The plan
of this paper is the following. In Sec.~\ref{sec:2}, we perform a
brief discussion about the existence of black hole remnant within
the noncommutative coordinate fluctuations at short distances
(noncommutative inspired Reissner-Nordstr\"{o}m solutions). We pay
special attention to study of Hawking temperature. In
Sec.~\ref{sec:3}, a detailed calculation of quantum tunneling near
the smeared quantum horizon by considering a new model of
noncommutativity is provided. The tunneling amplitude at which
charged massive particles tunnel across the event horizon is
computed and its applicability for the Reissner-Nordstr\"{o}m black
hole is discussed. And finally the paper is ended with summary
(Sec.~\ref{sec:4}).

\section{\label{sec:2}Noncommutative Reissner-Nordstr\"{o}m Black Hole}
There exist many formulations of noncommutative field theory based
on the Weyl-Wigner-Moyal $\ast$-product \cite{wey,wey.1,wey.2} that
lead to failure in resolving of some important problems, such as
Lorentz invariance breaking, loss of unitarity and UV divergences of
quantum field theory. But recently, Smailagic and Spallucci
\cite{sma,sma.1,sma.2,sma.3,sma.4} explained a fascinating model of
noncommutativity, the coordinate coherent states approach, that can
be free from the problems mentioned above. In this approach, a
point-like mass $M$, and a point-charge $Q$ instead of being quite
localized at a point, are described by a smeared structure
throughout a region of linear size $\sqrt{\theta}$. The approach we
adopt here is to look for a static, asymptotically flat, spherically
symmetric, minimal width, Gaussian distribution of mass and charge
whose noncommutative size is determined by the parameter
$\sqrt{\theta}$. To do this end, we shall model the mass and charge
distributions by a smeared delta function $\rho$ (see
\cite{nic1,nic2,nic2.1,ham1,ham2,riz,riz.1,riz.2,riz.3})
\begin{equation}\label{mat:1}  \Bigg\{
\begin{array}{ll}
\rho_{matt.}(r)={M\over
{(4\pi\theta)^{\frac{3}{2}}}}e^{-\frac{r^2}{4\theta}}\\
 \rho_{el.}(r)={Q\over {(4\pi \theta)^{\frac{3}{2}}}}
e^{-\frac{r^2}{4\theta}}.\\
\end{array}
\end{equation}
The line element which solves Einstein's equations in the presence
of smeared mass and charge sources can be obtained as
\begin{equation}
\label{mat:2}ds^2=-\bigg(1-\frac{2M_\theta}{r}+\frac{Q_\theta^2}{r^2}\bigg)dt^2+
\bigg(1-\frac{2M_\theta}{r}+\frac{Q_\theta^2}{r^2}\bigg)^{-1}dr^2+r^2
d\Omega^2,
\end{equation}
where $M_\theta$ and $Q_\theta$ are the smeared mass and charge
distributions respectively and can implicity be given in terms of
the lower incomplete Gamma function,
\begin{displaymath}
\left\{ \begin{array}{lll}
M_\theta=\frac{2M}{\sqrt{\pi}}\gamma\left(\frac{3}{2},\frac{r^2}{4\theta}\right)\\
Q_\theta=\frac{Q}{\sqrt{\pi}}\sqrt{\gamma^2\left(\frac{1}{2},\frac{r^2}{4\theta}\right)-
\frac{r}{\sqrt{2\theta}}\gamma\left(\frac{1}{2},\frac{r^2}{2\theta}\right)}\\
\gamma\left(\, \frac{a}{b},x\right)\equiv
 \int_0^x \frac{du}{u}\, u^{\frac{a}{b}} \, e^{-u}.\\
\end{array} \right.
\end{displaymath}
Throughout the paper, natural units are used with the following
definitions; $\hbar = c = G = k_B = 1$. In the limit
$\theta\rightarrow0$, the noncommutative (modified)
Reissner-Nordstr\"{o}m solution reduced to the commutative
(ordinary) case and $M_\theta\rightarrow M$, $Q_\theta\rightarrow Q$
as expected. The outer and inner horizons of this line element can
be found where $g_{00} ( r_{\theta\pm} ) = 0$, which are
respectively given by
\begin{equation}
\label{mat:3}r_{\theta\pm}(r_{\theta\pm})=M_\theta(r_{\theta\pm})\pm\sqrt{M_\theta^2(r_{\theta\pm})-Q_\theta^2(r_{\theta\pm})}.
\end{equation}
The noncommutative horizon radius versus the mass and charge can
approximately be calculated by setting $r_\pm=M\pm\sqrt{M^2-Q^2}$
into the lower incomplete Gamma function as
\begin{equation}
\label{mat:4}r_{\theta\pm}\equiv
r_{\theta\pm}(M,Q)=M_{\theta\pm}\pm\sqrt{M_{\theta\pm}^2-Q_{\theta\pm}^2},
\end{equation}
with
\begin{displaymath}
\left\{ \begin{array}{ll} M_{\theta\pm}\equiv
M_{\theta\pm}(M,Q)=M\left[{\cal{E}}\left(\frac{M\pm\sqrt{M^2-Q^2}}{2\sqrt{\theta}}\right)-\frac{M\pm\sqrt{M^2-Q^2}}{\sqrt{\pi
\theta}}\exp\left(-\frac{\left(M\pm\sqrt{M^2-Q^2}\right)^2}{4\theta}\right)\right] \\
Q_{\theta\pm}\equiv
Q_{\theta\pm}(M,Q)=Q\sqrt{{\cal{E}}^2\left(\frac{M\pm\sqrt{M^2-Q^2}}{2\sqrt{\theta}}\right)
 -\frac{M\pm\sqrt{M^2-Q^2}}{\sqrt{2\pi
\theta}}{\cal{E}}\left(\frac{M\pm\sqrt{M^2-Q^2}}{\sqrt{2\theta}}\right)},\\
\end{array} \right.
\end{displaymath}
where ${\cal{E}}(x)$ shows the {\it Gauss Error Function} defined as
$$ {\cal{E}}(x)\equiv \frac{2}{\sqrt{\pi}}\int_{0}^{x}e^{-t^2}dt.$$
For very large masses, the
${\cal{E}}\left(\frac{M\pm\sqrt{M^2-Q^2}}{2\sqrt{\theta}}\right)$
tends to unity and the exponential term will be reduced to zero and
one retrieves the classical Reissner-Nordstr\"{o}m horizons,
$r_{\theta\pm}\rightarrow r_\pm= M\pm\sqrt{M^2-Q^2}$.

\begin{figure}[htp]
\begin{center}
\includegraphics{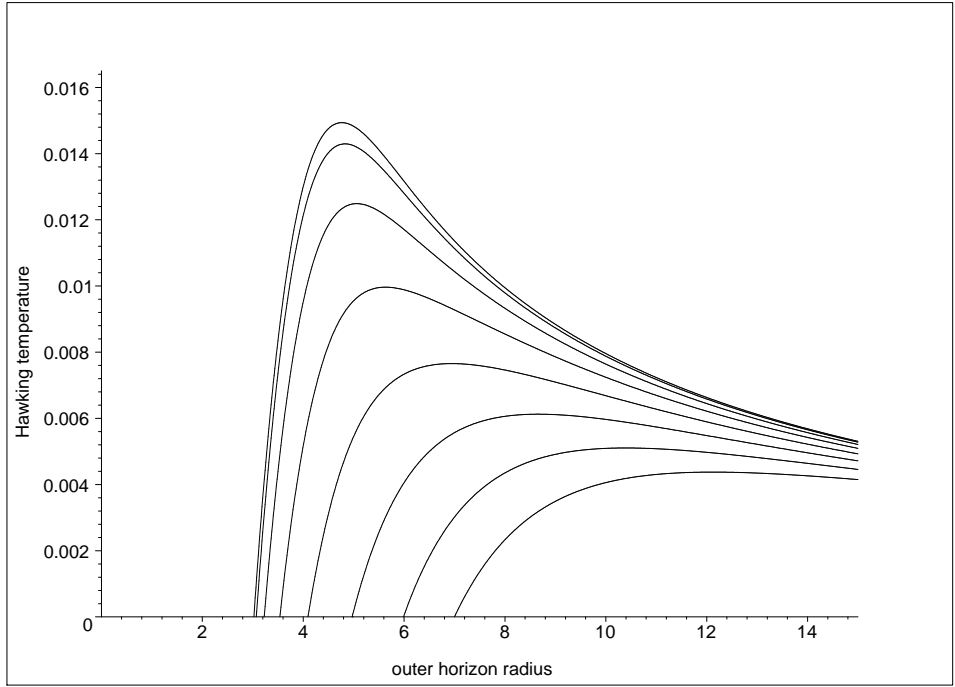}
\end{center}
\vspace{8.3 cm} \caption{\scriptsize { Black hole temperature,
$T_H\sqrt{\theta}$, as a function of
$\frac{r_{\theta+}}{\sqrt{\theta}}$ (the outer horizon radius), for
some values of $Q$. On the right-hand side of the figure, from top
to bottom, the curves correspond to $Q = 0,~ 1,~ 2,~ 3,~ 4,~ 5,~ 6$
and $7$, respectively. The top temperature reduces with growing $Q$.
The existence of a minimal non-zero mass and disappearance of
divergence are clear. In fact as a result of coordinate
noncommutativity the black hole temperature falls to zero at the
remnant mass.
 }}\label{fig:1}
\end{figure}

The radiating behavior of such modified Reissner-Nordstr\"{o}m black
hole can now be found to calculate its temperature as follows
$$T_H={1\over {4\pi}} {{dg_{00}}\over {dr}}\Big|_{r=r_{\theta+}} =
 \frac{1}{4\pi r_{\theta+}} \left[ {1 - \frac{r^3_{\theta+} \exp \left( -\frac{r^2_{\theta+}}{4\theta}\right)}{4\theta ^{\frac{3}{2}}
\gamma\left(\frac{3}{2},\frac{r_{\theta+}^2}{4\theta}\right)}}
\right]$$\begin{equation}\label{mat:5} -\frac{{Q^2 }}{{\pi^2
r^3_{\theta+} }} \left[
{\gamma^2\left(\frac{3}{2},\frac{r_{\theta+}^2}{4\theta}\right) +
\frac{r^3_{\theta+} \exp \left(
-\frac{r^2_{\theta+}}{4\theta}\right)\left(\gamma^2\left(\frac{1}{2},\frac{r_{\theta+}^2}{4\theta}\right)-
\frac{r_{\theta+}}{\sqrt{2\theta}}\gamma\left(\frac{1}{2},\frac{r_{\theta+}^2}{2\theta}\right)\right)}
{{16\,\theta ^{\frac{3}{2}}
\gamma\left(\frac{3}{2},\frac{r_{\theta+}^2}{4\theta}\right)}} }
\right].
\end{equation}
For the commutative case,
$\left(\frac{M+\sqrt{M^2-Q^2}}{2\sqrt{\theta}}\right)\rightarrow\infty$,
one recovers the classical Hawking temperature,
$T_H=\frac{\sqrt{M^2-Q^2}}{2\pi\left(M+\sqrt{M^2-Q^2}\right)^2}$.
The numerical calculation of such noncommutative Hawking temperature
as a function of $r_{\theta+}$, for some values of $Q$ is presented
in Fig.~\ref{fig:1}.
In this modified version, not only the Hawking temperature does not
diverge at all but also it reaches a maximum value before dropping
to absolute zero at a minimal non-zero mass, $M=M_0$, that black
hole shrinks to. In other words, Fig.~\ref{fig:1} shows that
coordinate noncommutativity yields to the existence of a minimal
non-zero mass which black hole can reduce to. Therefore, in the
noncommutative framework, black hole doesn't evaporate completely
and this leads to a Planck-sized remnant including {\it the
information}. So information might be preserved in this remnant.
However, it is not conceivable to date to give a clear answer to the
question of the black hole information paradox and this is
reasonable because there is no complete self-consistent quantum
theory of evaporating black holes (for reviews on resolving the
so-called {\it information loss paradox}, see
\cite{pre,pre.1,pre.2,pre.3,pre.4,pre.5}).

In this situation, we should note that our calculation to obtain the
Eq.~(\ref{mat:5}) is accurate and no approximation has been made. If
we want to acquire the simple \,$r_{\theta\pm}$-dependent form of
the noncommutative Hawking temperature, then it can be approximated
as follows
\begin{equation}
\label{mat:5.1}T_H=\frac{\kappa(M,Q)}{2\pi}\approx\frac{1}{4\pi}\frac{r_{\theta+}-r_{\theta-}}{r^2_{\theta+}},
\end{equation}
where $\kappa(M,Q)$ is the noncommutative surface gravity at the
horizon and is given by
\begin{equation}
\label{mat:5.2}\kappa(M,Q)\approx
\frac{r_{\theta+}-r_{\theta-}}{2r^2_{\theta+}}.
\end{equation}
In Sec.~\ref{sec:3}, we will use this approximate expression to
compute the tunneling probability when the first law of black hole
thermodynamics is applied (see Eq.~(\ref{mat:23})).

\section{\label{sec:3} Parikh-Wilczek Tunneling as Charged Massive Particles}
We are now ready to discuss the quantum tunneling process in the
noncommutative framework. To describe this procedure, where a
particle moves in dynamical geometry and pass through the horizon
without singularity on the path we should use the coordinates
systems that, unlike Reissner-Nordstr\"{o}m coordinates, are not
singular at the horizon (the outer horizon). A particularly
convenient choice is Painlev\'{e} coordinate \cite{pai} which is
obtained by definition of a new noncommutative time coordinate,
\begin{equation}
\label{mat:6}dt=dt_r+\frac{r\sqrt{2M_{\theta+}r-Q_{\theta+}^2}}{r^2-2M_{\theta+}r+Q_{\theta+}^2}dr=dt_r+dt_{syn},
\end{equation}
where $t_r$ is the Reissner time coordinate, and
\begin{equation}
\label{mat:7}dt_{syn}=-\frac{g_{01}}{g_{00}}dr.
\end{equation}
Note that only the Reissner time coordinate is transformed. Both the
radial coordinate and angular coordinates remain the same. The
noncommutative Painlev\'{e} metric now immediately reads
$$ds^2 =
g_{00}dt^2+2g_{01}dtdr+g_{11}dr^2+g_{22}d\vartheta^2+g_{33}d\varphi^2$$
\begin{equation}\label{mat:8}=-\bigg(1 -
\frac{2M_{\theta+}}{r}+\frac{Q_{\theta+}^2}{r^2} \bigg) dt^2 +
2\sqrt{\frac{2M_{\theta+}}{r}-\frac{Q_{\theta+}^2}{r^2}} dtdr + dr^2
+ r^2( d\vartheta^2+\sin^2\vartheta d\varphi^2).
\end{equation}
It should be stressed here that the Eq.~(\ref{mat:7}), in accord
with Landau's theory of the synchronization of clocks \cite{lan},
allows us to synchronize clocks in any infinitesimal radial
positions of space ($d\vartheta=d\varphi=0$). Since the tunneling
phenomena through the quantum horizon i.e. {\it the barrier} is an
instantaneous procedure it is important to consider Landau's theory
of the coordinate clock synchronization in the tunneling process.
The mechanism for tunneling through the quantum horizon is that
particle anti-particle pair is created at the event horizon. So, we
have two events that occur simultaneously; one event is
anti-particle and tunnels into the barrier but the other particle
tunnels out the barrier. In fact, the Eq.~(\ref{mat:7}) mentions the
difference of coordinate times for these two simultaneous events
occurring at infinitely adjacent radial positions. Furthermore, the
noncommutative Painlev\'{e}-Reissner-Nordstr\"{o}m metric exhibits
the stationary, non-static, and neither coordinate singularity nor
intrinsic singularity.

To obtain the radial geodesics of the charged massive particles'
tunneling across the potential barrier which are different with the
uncharged massless \cite{ham1} and also the uncharged massive
\cite{ham2} ones, we follow the same noncommutative method in this
work but now extended to the case of charged massive particles.
According to the non-relativistic quantum theory, de Broglie's
hypothesis and the WKB approximation, it can be easily proved that
the treatment of the massive particle's tunneling as a massive shell
is approximately derived by the phase velocity $v_p$ of the de
Broglie s-wave whose the relationship between phase velocity $v_p$
and group velocity $v_g$ is given by \cite{zha,zha.1,zha.2}
\begin{equation}
\label{mat:9}v_p=\dot{r}=\frac{1}{2}v_g,
\end{equation}
overdot abbreviates $\frac{d}{dt}$. In the case of
$d\vartheta=d\varphi=0$, according to the relation (\ref{mat:7}),
the group velocity is
\begin{equation}
\label{mat:10}v_g=-\frac{g_{00}}{g_{01}}.
\end{equation}
Thus, the outgoing motion of the massive particles take the form
\begin{equation}
\label{mat:11}\dot{r}=-\frac{g_{00}}{2g_{01}}=
\frac{r^2-2M_{\theta+}r+Q_{\theta+}^2}{2r\sqrt{2M_{\theta+}r-Q_{\theta+}^2}}.
\end{equation}
If we suppose that $t$ increases towards the future then the above
equations will be modified by the particle's self-gravitation effect
\cite{kra,kra.1} (see also \cite{sha,sha.1}) . To assure the
conservation of energy and electric charge, we fix the total ADM
mass ($M$) and electric charge ($Q$) of the spacetime and permit the
hole mass and its charge to fluctuate. In other words, we should
replace $M$ by $M-E$ and $Q$ by $Q-q$ both in the Eqs.~(\ref{mat:8})
and (\ref{mat:11}), because the response of the background geometry
is taken into account by an emitted quantum of energy $E$ with
electric charge $q$. Thus, when a charged particle tunnels out, the
black holes's mass and also electric charge will change for the
conservation of energy and charge.

\begin{figure}[htp]
\begin{center}
\includegraphics{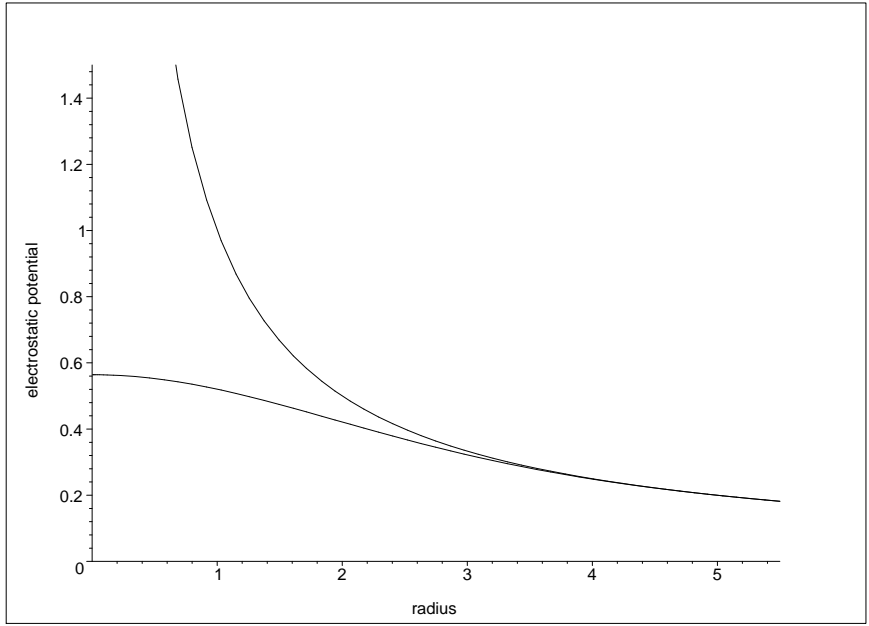}
\end{center}
\vspace{8 cm} \caption{\scriptsize {The electrostatic potential (per
unit charge) versus the radius. The upper curve is the ordinary
Coulomb potential and the lower curve is the result of
noncommutativity. The regular behavior of the the Coulomb-like
potential at the origin can be easily seen from the lower curve.}}
\label{fig:2}
\end{figure}

In order to consider the effect of the electromagnetic field, it is
necessary to take into account Maxwell gravity system comprises of
the black hole and the electromagnetic field outside the hole. The
lagrangian function of the Maxwell gravity system should be written
as
\begin{equation}
\label{mat:12}L=L_{matt.}+L_{el.},
\end{equation}
where $L_{el.}=-\frac{1}{4}F_{\mu\nu}F^{\mu\nu}$ is the lagrangian
function, while the Maxwell field $F^{\mu\nu}$ must take on the form
\begin{equation}
\label{mat:13}F^{\mu\nu} = \delta^{0[\, \mu\,\vert}
\delta^{r\,\vert\, \nu \,]}\, E(r\,;\,\theta)=\partial^\mu
\phi^\nu-\partial^\nu \phi^\mu.
\end{equation}
Studying the behavior of Coulomb-like field within the
noncommutativity framework have already been investigated in
Refs.~\cite{nic2,nic2.1}. The Electric field $E(r\,;\,\theta)$ is
found by solving the following Maxwell equations with a
Gaussian-profile of smearing-charge source along the time direction:
\begin{equation}
\label{mat:14} \frac{1}{\sqrt{-g}}\, \partial_\mu\,\left(\,
\sqrt{-g}\, F^{\mu\nu}\, \right) = \rho_{el.}\,\delta^\nu_0,
\end{equation}
which is given as
\begin{equation}
\label{mat:15}E(r\,;\,\theta)=\frac{2Q}{\sqrt\pi\, r^2}\,
\gamma\left(\frac{3}{2}, \frac{r^2}{4\theta}\right).
\end{equation}
The regular behavior of the Coulomb-like field at the origin is
clear. In the limit of \, $\theta\rightarrow0$, the lower incomplete
Gamma function reduces to the complete Gamma function
$\Gamma(\frac{3}{2})$, and one recovers the ordinary Coulomb field.
The last equality of the Eq.~(\ref{mat:13}) defines $F^{\mu\nu}$ in
terms of the 4-potential and corresponds to the generalized
coordinates $\phi_\mu = (\phi, 0, 0, 0)$ \cite{mak}, where the
Coulomb-like potential $\phi$ is the only non-zero component of the
electromagnetic potential $\phi_\mu$ and can be obtained as follows:
$$\phi\equiv\phi(r\,;\,\theta)=$$\begin{equation}
\label{mat:16}-\frac{Q}{2\sqrt{\pi\theta}}\Bigg[\frac{r^2}{12\theta}\,\,
\,_3F_3\left(1, 1, \frac{3}{2} \,\, ; \,\, 2, 2, \frac{5}{2} \,\, ;
\,\,
-\frac{r^2}{4\theta}\right)-\Gamma\left(0,\frac{r^2}{4\theta}\right)-2\ln\bigg(\frac{r^2}{\theta}\bigg)
 +2\ln2+2-\gamma\Bigg].
\end{equation}
In the above equation,\, $\,_pF_q$ shows the {\it hypergeometric
series} defined in terms of the {\it pochhammer symbol} as follows:
\begin{displaymath}
\,_pF_q\left(a_1, \ldots , a_p\,\,;\,\,b_1, \ldots , b_q\,\,;
z\right)=\sum_{n=0}^{\infty}\frac{z^n\bigg(\prod_{i=1}^{p}
pochhammer(a_i,n)\bigg)}{n!\bigg(\prod_{j=1}^{q}
pochhammer(b_j,n)\bigg)}.
\end{displaymath}
where, for example, $pochhammer(a_i,n)=a_i(a_i+1)\ldots(a_i+n-1)$.
The second term, $\Gamma\left(0,\frac{r^2}{4\theta}\right)$, is the
upper incomplete Gamma function which is defined as
\begin{displaymath}
\Gamma\left(0,\frac{r^2}{4\theta}\right)=\int_{\frac{r^2}{4\theta}}^\infty\frac{e^{-t}}{t}dt.
\end{displaymath}
The last term of the Eq.~(\ref{mat:16}), $\gamma$, is the {\it
Euler's constant} that is approximately equal to 0.577215. To have a
clear depiction, we have computed the numerical results of both the
noncommutative Coulomb-like potential and commutative case which are
shown in Fig.~\ref{fig:2}.
When a charged particle passes through the event horizon, the whole
system will transit from one state to another. In order to remove
the freedom equivalent to $\phi$, due to the fact that it is an
ignorable coordinate, the action is then found to be
\begin{equation}
\label{mat:17}I=\int_{t_{in}}^{t_{out}}\left(L-P_{\phi}\dot{\phi}\right)dt,
\end{equation}
where $P_{\phi}$ is the canonical momentum conjugate to $\phi$.
Since the characteristic wavelength of the radiation is always
haphazardly small near the horizon due to the infinite blue-shift
there, so that the wave-number reaches infinity and the WKB
approximation is reliable close to the horizon. In the WKB
approximation, the probability of tunneling or emission rate for the
classically forbidden region as a function of the imaginary part of
the particle's action at stationary phase would take the form
{\footnote {We should stress that there is another opinion on
utilizing the expression (\ref{mat:18}). There is a problem here
recognized as "factor $2$ problem" \cite{akh,akh.1}. Lately, some
authors (see \cite{cho,pil} and references therein) have stated that
the expression (\ref{mat:18}) is not invariant under canonical
transformations however the same formula with a factor of $1/2$ in
the exponent is canonically invariant. This procedure yields a
temperature higher than the Hawking temperature by a factor of $2$.
In Refs.~\cite{akh2,akh2.1}, a resolution to this problem was given
in terms of an overlooked temporal contribution to the tunneling
amplitude. When one includes this temporal contribution one gets
exact the correct temperature and exactly when one uses the
canonically invariant tunneling amplitude.}}
\begin{equation}
\label{mat:18}\Gamma\sim\exp(-2\textmd{Im}\, I).
\end{equation}
To calculate the imaginary part of the action we consider a
spherical shell to consist of components of the charged massive
particles each of which travels on a radial timelike geodesic, so
that we will use these radial timelike geodesics like an s-wave
outgoing positive energy particle which pass through the horizon
outwards from $r_{in}$ to $r_{out}$ to compute the $\textmd{Im}\,
I$, as follows
\begin{equation}
\label{mat:19}\textmd{Im}\,
I=\textmd{Im}\int_{r_{in}}^{r_{out}}\left(P_r-\frac{P_{\phi}\dot{\phi}}{\dot{r}}\right)dr
=\textmd{Im}\int_{r_{in}}^{r_{out}}\left[\int_{(0,0)}^{(P_r,P_{\phi})}dP'_r-\frac{\dot{\phi}}{\dot{r}}dP'_{\phi}\right]dr,
\end{equation}
we now alter the integral variables by using Hamilton's equations
\begin{equation}\label{mat:20}  \Bigg\{
\begin{array}{ll}
\dot{r}=\frac{dH}{dP_r}\big|_{(r;\,\phi,\,P_{\phi})}=\frac{d(M-E)}{dP_r}=-\frac{dE}{dP_r}\\
 \dot{\phi}=\frac{dH}{dP_{\phi}}\big|_{(\phi;\,r,\,P_r)}=\phi(Q-q)\,\,\frac{dq}{dP_{\phi}}.\\
\end{array}
\end{equation}
Hence, the imaginary part of the action gives the following
expression:
\begin{equation}
\label{mat:21}\textmd{Im}\, I
=-\textmd{Im}\int_{r_{in}}^{r_{out}}\left[\int_{(0,0)}^{(E,q)}\frac{dE'}{\dot{r}}+\frac{\phi(Q-q')}{\dot{r}}dq'\right]dr.
\end{equation}
Substituting the expression of $\dot{r}$ from (\ref{mat:11}) into
(\ref{mat:21}), under the condition that the self-gravitation effect
of the particle itself is included, we have,
\begin{equation}
\label{mat:22}\textmd{Im}\, I
=-\textmd{Im}\int_{r_{in}}^{r_{out}}\left[\int_{(0,0)}^{(E,q)}\frac{2r\sqrt{2M'_{\theta+}r
-Q'^2_{\theta+}}}{r^2-2M'_{\theta+}r+Q'^2_{\theta+}}dE'+\frac{2r\,\phi(Q-q')\,\sqrt{2M'_{\theta+}r
-Q'^2_{\theta+}}}{r^2-2M'_{\theta+}r+Q'^2_{\theta+}}dq'\right]dr,
\end{equation}
where
\begin{displaymath}
\left\{ \begin{array}{ll}
M'_{\theta+}\equiv M_{\theta+}(M-E',Q-q') \\
Q'_{\theta+}\equiv Q_{\theta+}(M-E',Q-q').
\end{array} \right.
\end{displaymath}
The $r$ integral has a pole at the outer horizon where lies along
the line of integration. This integral can be done first by
deforming the contour and it yields to ($-\pi i$) times the residue.
Note that we require $r_{in}>r_{out}$ where:
\begin{displaymath}
\left\{ \begin{array}{ll}
 r_{in}=\frac{2M}{\sqrt{\pi}}\gamma\Big(\frac{3}{2},\frac{r_{in}^2}{4\theta}\Big)
 +\sqrt{\frac{4M^2}{\pi}\gamma^2\Big(\frac{3}{2},\frac{r_{in}^2}{4\theta}\Big)-
 \frac{Q^2}{\pi}\left(\gamma^2\Big(\frac{1}{2},\frac{r_{in}^2}{4\theta}\Big)-\frac{r_{in}}{\sqrt{2\theta}}
 \gamma\Big(\frac{1}{2},\frac{r_{in}^2}{2\theta}\Big)\right)}\\
 r_{out}=\frac{2(M-E)}{\sqrt{\pi}}\gamma\Big(\frac{3}{2},\frac{r_{out}^2}{4\theta}\Big)
 +\sqrt{\frac{4(M-E)^2}{\pi}\gamma^2\Big(\frac{3}{2},\frac{r_{out}^2}{4\theta}\Big)-
 \frac{(Q-q)^2}{\pi}\left(\gamma^2\Big(\frac{1}{2},\frac{r_{out}^2}{4\theta}\Big)-\frac{r_{out}}{\sqrt{2\theta}}
 \gamma\Big(\frac{1}{2},\frac{r_{out}^2}{2\theta}\Big)\right)}\\
\end{array} \right.
\end{displaymath}
Thus,
\begin{equation}
\label{mat:23}\textmd{Im}\,I=2\pi\left[\int_{(0,0)}^{(E,q)}
\frac{r'^2_{\theta+}}{r'_{\theta+}-r'_{\theta-}}\,dE'+\frac{r'^2_{\theta+}\,\,\phi(Q-q')}{r'_{\theta+}-r'_{\theta-}}\,dq'\right]
=\pi\left[\int_{(0,0)}^{(E,q)}\frac{dE'}{\kappa'}-\frac{V'}{\kappa'}dq'\right],
\end{equation}
where
\begin{displaymath}
\left\{ \begin{array}{lll}
r'_{\theta\pm}\equiv r_{\theta\pm}(M-E',Q-q') \\
V'=-\phi(Q-q')\Big|_{r=r'_{\theta+}} \\
\kappa'\equiv\kappa(M-E',Q-q')=
\frac{r'_{\theta+}-r'_{\theta-}}{2r'^2_{\theta+}}.
\end{array} \right.
\end{displaymath}
In the above expressions, $V'$ and $\kappa'$, respectively, are the
electro-potential on the event horizon and the horizon surface
gravity in which self-gravitations are comprised. Here, utilizing
the first low of black hole thermodynamics,
$dM=\frac{\kappa}{2\pi}dS+VdQ$, one can find the imaginary part of
the action as
\cite{kes,mas,med,med.1,med.2,ban,ban.1,ban.2,ban2,ban2.1}
\begin{equation}
\label{mat:24}\textmd{Im}\,I=-\frac{1}{2}\int_{S_{NC}(M,Q)}^{S_{NC}(M-E,Q-q)}dS=-\frac{1}{2}\Delta
S_{NC}(M,Q,E),
\end{equation}
where $\Delta S_{NC}(M,Q,E)$ is the difference in noncommutative
black hole entropies before and after emission. Since, both particle
and anti-particle (which corresponds to a time reversed situation
and it can be seen that as backward in time by replacing\,
$\sqrt{\frac{2M_{\theta+}}{r}-\frac{Q_{\theta+}^2}{r^2}}$, \,by\,
$-\sqrt{\frac{2M_{\theta+}}{r}-\frac{Q_{\theta+}^2}{r^2}}$,\, in the
metric (\ref{mat:8})) anticipate in the emission rate for the
Hawking process via tunneling with same amounts, therefore we should
have to add their amplitudes first and then to square it to obtain
the emission probability,
\begin{equation}
\label{mat:25}\Gamma\sim\exp(-2\textmd{Im}\, I)\sim\exp\left(\Delta
S_{NC}(M,Q,E)\right)=\exp[S_{NC}(M-E,Q-q)-S_{NC}(M,Q)].
\end{equation}
Hawking radiation as tunneling was also investigated in the context
of black holes in string theory \cite{kes}, and it was exhibited
that the emission rates in the high energy corresponds to a
difference between counting of states in the microcanonical and
canonical ensembles. Thus at higher energies the emission spectrum
cannot be precisely thermal due to the fact that the high energy
corrections arise from the physics of energy and charge conservation
with noncommutativity corrections. In fact, the emission rate
(\ref{mat:25}) deviates from the pure thermal emission but is
consistent with an underlying unitary quantum theory and support the
conservation of information \cite{par2}. The question which arises
here is the possible dependencies between different modes of
radiation during the evaporation \cite{ham3,ham3.1,ham4} and then
the time-evolution of these possible correlations which needs
further investigation and probably shed more light on information
loss problem. This problem is currently under investigation.

\section{\label{sec:4}Summary}
The generalization of the standard Hawking radiation via tunneling
through the event horizon based on the solution of the
Eq.~(\ref{mat:18}) within the context of coordinate coherent state
noncommutativity has been studied and then the new corrections of
the emission rate based on noncommutative framework has been
achieved. To describe the noncommutative behavior of an
electro-gravitational system, we have extended the Parikh-Wilczek
tunnelling process to calculate the tunneling probability of a
charged massive particle from the Reissner-Nordstr\"{o}m black hole
within the framework of noncommutative quasi coordinates. Studying
its behavior shows that, as expected, the emission rate is
consistent with the unitary theory and satisfies the first law of
black hole thermodynamics.

\end{document}